\documentclass[%
	aps,
	prd,
	reprint,
	superscriptaddress,
	floatfix,
	nofootinbib,
	noamsmath,
	noamssymb
]
{revtex4-2}

\usepackage[T1]{fontenc}
\usepackage[utf8]{inputenc}
\usepackage{graphicx}
\usepackage{xcolor}
\usepackage{mathtools}
\usepackage{amssymb}
\usepackage{lmodern}
\usepackage{bm}
\usepackage{siunitx}
\usepackage[final]{microtype}
\usepackage{hyperref}

\newcommand{\mytitle}{Critical endpoint of QCD in a finite volume}

\newcommand{\JLU}{%
	Institut f\"{u}r Theoretische Physik, %
	Justus-Liebig-Universit\"{a}t Gie\ss{}en, %
	35392 Gie\ss{}en, %
	Germany%
}

\newcommand{\HFHF}{%
	Helmholtz Forschungsakademie Hessen f\"{u}r FAIR (HFHF), %
	GSI Helmholtzzentrum f\"{u}r Schwerionenforschung, %
	Campus Gie\ss{}en, %
	35392 Gie\ss{}en, %
	Germany%
}

\hypersetup{%
	pdftitle = {\mytitle},
	pdfauthor = {Julian Bernhardt, Christian S. Fischer, Philipp Isserstedt, Bernd-Jochen Schaefer},
	bookmarksopen = true,
	colorlinks = true,
	linkcolor = green!50!black,
	citecolor = blue!50!black,
	urlcolor = blue!50!black
}

\graphicspath{{./Graphics/}}

\DeclareSIUnit{\MeV}{\mega\electronvolt}
\DeclareSIUnit{\GeV}{\giga\electronvolt}
\DeclareSIUnit{\fm}{\femto\meter}

\DeclarePairedDelimiter{\abs}{\lvert}{\rvert}
\DeclarePairedDelimiter{\expval}{\langle}{\rangle}
\DeclareMathOperator{\Tr}{Tr}
\DeclareMathOperator*{\argmax}{argmax}

\renewcommand*{\vec}[1]{\bm{#1}}
\newcommand*{\+}{\hspace*{.08335em}}
\newcommand*{\dd}{\mathrm{d}}
\newcommand*{\ii}{\mathrm{i}}
\newcommand*{\bbZ}{\mathbb{Z}}
\newcommand*{\bbR}{\mathbb{R}}
\newcommand*{\calL}{\mathcal{L}}

\newcommand*{\upu}{\textup{u}}
\newcommand*{\upd}{\textup{d}}
\newcommand*{\ups}{\textup{s}}
\newcommand*{\upB}{\textup{B}}
\newcommand*{\upvol}{\textup{vol}}

\newcommand*{\upzero}{\textup{zero}}
\newcommand*{\upmin}{\textup{min}}
\newcommand*{\PBCs}{\textup{PBC}^{\ast}}
\newcommand*{\ZT}{Z_{\textup{T}}}
\newcommand*{\ZL}{Z_{\textup{L}}}
\newcommand*{\ProjT}[1]{P_{#1}^\textup{T}}
\newcommand*{\ProjL}[1]{P_{#1}^\textup{L}}
\newcommand*{\OO}{\operatorname{O}}
\newcommand*{\Tc}{T_{\textup{c}}}

\DeclareMathOperator*{\sumint}{%
	\mathchoice{%
		\ooalign{%
			$\displaystyle\sum$\cr\hidewidth$\displaystyle\int$\hidewidth\cr%
		}%
	}{%
		\ooalign{%
			\raisebox{0.14\height}{%
				\scalebox{0.7}{%
					$\textstyle\sum$%
				}
			}%
			\cr\hidewidth$\textstyle\int$\hidewidth\cr%
		}%
	}{%
		\ooalign{%
			\raisebox{0.2\height}{%
				\scalebox{0.6}{%
					$\textstyle\sum$%
				}
			}%
			\cr$\scriptstyle\int$\cr%
		}%
	}{%
		\ooalign{%
			\raisebox{0.2\height}{%
				\scalebox{0.6}{%
					$\textstyle\sum$%
				}
			}%
			\cr$\scriptstyle\int$\cr%
		}%
	}%
}

\begin{document}

\title{\mytitle}

\author{Julian Bernhardt}
\email{julian.bernhardt@physik.uni-giessen.de}
\affiliation{\JLU}
\affiliation{\HFHF}

\author{Christian S.~Fischer}
\email{christian.fischer@theo.physik.uni-giessen.de}
\affiliation{\JLU}
\affiliation{\HFHF}

\author{Philipp Isserstedt}
\email{philipp.isserstedt@physik.uni-giessen.de}
\affiliation{\JLU}
\affiliation{\HFHF}

\author{Bernd-Jochen Schaefer}
\email{bernd-jochen.schaefer@theo.physik.uni-giessen.de}
\affiliation{\JLU}
\affiliation{\HFHF}

\begin{abstract}
We investigate the impact of finite volume and the corresponding restrictions
on long-range correlations on the location of the critical endpoint in the QCD
phase  diagram. To this end, we employ a sophisticated combination of lattice
Yang--Mills theory and a (truncated) version of Dyson--Schwinger equations in
Landau gauge for $2 + 1$ quark flavors that has been studied extensively
in the past. In the infinite-volume limit, this system predicts a critical
endpoint at moderate temperature and large chemical potential. We study this
system at small and intermediate volumes and determine the dependence of the
location of the critical endpoint on the boundary conditions and the volume of a
three-dimensional cube with edge length $L$. We demonstrate that noticeable
volume effects of more than five percent occur only for $L \lesssim \SI{5}{\fm}$
and that volumes as large as $L^3 \gtrsim (\SI{8}{\fm})^3$ are very close to
the infinite-volume limit.
\end{abstract}

\maketitle

\section{\label{introduction}%
	Introduction
}

There are a number of reasons why finite-volume studies of the location and
properties of a putative chiral critical endpoint (CEP) in the phase diagram of
QCD are interesting. First, the search of such a CEP in heavy-ion collisions
produces an initial fireball of finite extent with typical scales of several
femtometers in each direction. Volume effects on important observables such as
fluctuations of conserved charges may be considerable and need to be taken into
account \cite{Luo:2017faz,Bzdak:2019pkr}. Second, the location of all phase
boundaries including the crossover at zero chemical potential, the CEP, and the
chiral first-order transition will certainly become volume dependent for small
enough volumes. At small chemical potential, these are accessible by various
extrapolation methods applied to lattice QCD. The corresponding simulations are
carried out at finite boxes with (anti)periodic boundary conditions, and a
thorough understanding of the volume dependence of the results is mandatory.
Third, effects due to changes in the volume are not only annoying artifacts but
interesting in themselves because they serve to probe the reaction of a
physical system on one of its external parameters. This is particularly
interesting when it comes to small quark masses and important questions on the
properties of the upper and lower left (chiral) corners in the Columbia plot.

Theoretical studies of finite-volume effects on the QCD phase diagram have been
carried out in a number of approaches besides lattice QCD. Model studies in the
Nambu--Jona-Lasinio or the quark-meson model and their Polyakov-loop enhanced
versions serve to highlight important general aspects of small-volume physics;
see, e.g., Refs.~\cite{Braun:2011iz,Skokov:2012ds,Bhattacharyya:2012rp,
Bhattacharyya:2014uxa} and Ref.~\cite{Klein:2017shl} for a review.
In particular, the renormalization-group treatment of these models allows to
study the reaction of fluctuations on changes of volume \cite{Braun:2011iz,
Tripolt:2013zfa,Juricic:2016tpt}. Functional approaches to QCD via
Dyson--Schwinger equations (DSEs) or the functional renormalization group (FRG)
have the additional benefit of treating the Yang--Mills sector explicitly,
which allows for quantitative studies on the location of the CEP
\cite{Fischer:2014ata,Isserstedt:2019pgx,Fu:2019hdw,Gao:2020qsj,Gao:2020fbl}.

In this work, we employ the DSE framework of Ref.~\cite{Isserstedt:2019pgx} and
treat the system in a finite three-dimensional box of equal edge lengths $L$.
We contrast periodic and antiperiodic boundary conditions and vary the box
size between $L = \SI{3}{\fm}$ and $L = \SI{8}{\fm}$. We trace the location of
the CEP in the QCD phase diagram for various volumes and determine the box size
necessary to approach the infinite-volume results. Furthermore, we discuss
volume effects on the curvature of the crossover line at small chemical
potential and compare with lattice QCD.

The paper is organized as follows. In Sec.~\ref{framework}, we discuss all
technical details of our framework. We briefly summarize the truncation scheme
for the DSEs and explain our implementation of finite volume. In particular, we
explain a method to get rid of cubic artifacts at large momenta in the
ultraviolet (UV) that drastically improves the approach to the infinite-volume
limit. In Sec.~\ref{results}, we discuss our results for the volume dependence
of the chiral crossover, its curvature, and the CEP and compare UV improved
with unimproved calculations. Finally, we summarize in Sec.~\ref{summary}.

\section{\label{framework}%
	Framework
}

\subsection{\label{framework:prop_fv}%
	In-medium propagators and finite volume
}

The quantities of interest in this work are the dressed quark and gluon
propagators $S_{f}$ and $D_{\nu\sigma}$, respectively, where $f$ labels the
quark flavor. In Landau gauge and at nonzero temperature $T$ and quark chemical
potential $\mu_{f}$, they are given by
\begin{align}
	\label{eq:quark_propagator}
	S_{f}^{-1}(p)
	&=
	\ii \+ \omega_{n} \gamma_{4} \+ C_{f}(p)
	+
	\ii \+ \vec{\gamma} \cdot \vec{p} \+ A_{f}(p)
	+
	B_{f}(p) \, ,
	\displaybreak[0]
	\\[0.5em]
	\label{eq:gluon_propagator}
	D_{\nu\sigma}(p)
	&=
	\frac{\ZT(p)}{p^{2}} \+ \ProjT{\nu\sigma}(p)
	+
	\frac{\ZL(p)}{p^{2}} \+ \ProjL{\nu\sigma}(p) \, ,
\end{align}
where $p = (\omega_{n}, \vec{p})$ is the four-momentum with Matsubara
frequencies $\omega_{n}$, $n \in \bbZ$.\+%
\footnote{We work in four-dimensional Euclidean space-time with
positive metric signature (++++). The Euclidean gamma matrices obey
$\{ \gamma_{\nu}, \gamma_{\sigma} \} = 2 \+ \delta_{\nu\sigma}$.}
An additional quark tensor structure $\gamma_{4} \+ \vec{\gamma} \cdot \vec{p}$
is in principle possible but negligible \cite{Contant:2017gtz}. The dressing
functions $C_{f}$, $A_{f}$, and $B_{f}$ of the quark and $\ZT$ and $\ZL$
of the gluon have a nontrivial momentum dependence and contain all
nonperturbative information. Furthermore, they depend implicitly on temperature
and quark chemical potential. The structure of Eqs.~\eqref{eq:quark_propagator}
and \eqref{eq:gluon_propagator} reflects the breaking of the $\OO(4)$ symmetry
in vacuum down to $\OO(3)$ at nonzero temperature due to the coupling to the
heat bath.\+%
\footnote{We use the customary choice of $(1, \vec{0})$ for the
heat-bath vector, i.e., we are in the rest frame of the medium.}
In particular, the gluon splits into a part transversal (T) and longitudinal
(L) with respect to the heat bath. The corresponding projectors read
\begin{align}
	\label{eq:projector_T}
	\ProjT{\nu\sigma}(p)
	&=
	(1 - \delta_{4\nu}) \+
	(1 - \delta_{4\sigma})
	\left(
		\delta_{\nu\sigma} - \frac{p_{\nu} \+ p_{\sigma}}{\vec{p}^{2}}
	\right) ,
	\\[0.5em]
	\label{eq:projector_L}
	\ProjL{\nu\sigma}(p)
	&=
	\delta_{\nu\sigma}
	-
	\frac{p_{\nu} \+ p_{\sigma}}{p^{2}}
	-
	\ProjT{\nu\sigma}(p) \, .
\end{align}
The fermionic Matsubara frequencies include the chemical potential,
$\omega_{n} = (2 \+ n + 1) \+ \pi T + \ii \+ \mu_{f}$, and the bosonic
frequencies read $\omega_{n} = 2 \+ n \+ \pi T$.

\subsubsection{\label{framework:prop_fv:pure_torus}%
	Finite volume without UV improvement
}

The introduction of a finite, uniform, three-dimensional spatial volume with
(anti)periodic boundary conditions has similarities with the introduction of
nonzero temperature. Within the Matsubara formalism, nonzero temperature is
implemented in the QCD action by compactifying the imaginary-time integration
to the finite interval $[0, 1 / \+ T]$, i.e.,
\begin{equation}
	\label{eq:action_nonzero_T}
	\int \dd^{4} x \, \calL
	\to
	\int_{0}^{1 / \+ T} \dd \tau \,
	\int_{\bbR^{3}} \dd^{3} x \, \calL \, ,
\end{equation}
where $\calL$ is the QCD Lagrangian. Restricting this action further to a finite
spatial cube with edge length $L$ amounts to the additional replacement
\begin{equation}
	\label{eq:action_finite_volume}
	\int_{\bbR^{3}} \dd^{3} x \, \calL
	\to
	\int_{[0, L]^3} \dd^{3} x \, \calL \, .
\end{equation}
As a consequence, in momentum space, the spatial three-momentum becomes discrete
in addition to the already discrete Matsubara frequencies $\omega_{n}$ in
(imaginary) time direction. Hence, each generic momentum integral with kernel
$K$ is replaced by a corresponding sum over spatial modes:
\begin{equation}
	\label{eq:integral_to_sum}
	\int \frac{\dd^{3} q}{(2 \pi)^{3}} \, K(\vec{q})
	\to
	\frac{1}{L^{3}} \sum_{\vec{z} \in \bbZ^{3}} K(\vec{q}_{\vec{z}})
\end{equation}
with $\vec{q}_{\vec{z}} = \sum_{i=1}^{3} \omega_{z_{i}}^{L} \vec{e}_{i}$ and
the Cartesian basis $\{ \vec{e}_{1}, \vec{e}_{2}, \vec{e}_{3} \}$ that spans
the three-dimensional momentum space. While the Matsubara frequencies in
temporal direction are fixed by the spin-statistics theorem---odd (even)
multiples of $\pi T$ for quarks (gluons)---there is no similar constraint in
spatial direction. Although it turns out that kinematics necessitates periodic
spatial boundary conditions for the gluon, the quarks are completely free. In
this work, we will choose periodic spatial boundary conditions (PBC) or
antiperiodic spatial boundary conditions (ABC) for the quarks according to
\begin{equation}
	\label{eq:matsubara_spatial}
	\omega_{n}^{L}
	=
	\begin{cases}
		2 \+ n \+ \pi / L & \text{for PBC} \, ,
		\\[0.5em]
		(2 \+ n + 1) \+ \pi / L & \text{for ABC} \, ,
	\end{cases}
\end{equation}
where $n \in \mathbb{Z}$. Note that PBC contain a zero mode
$\vec{q}_{\+\upzero} = \vec{0}$ corresponding to the lowest possible momentum.
Because its inclusion in both the  quark self-energy and the quark-loop
contribution to the gluon self-energy requires special attention, we will
discuss it in more detail below. Moreover, in order to compare with other
works, we also perform calculations where the PBC zero mode is omitted; we
refer to this setup as $\PBCs$.

\begin{figure}[t]
	\centering%
	\includegraphics[scale=1.0]{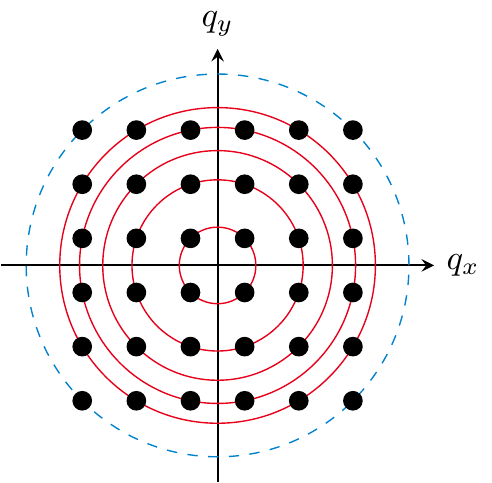}%
	\caption{\label{fig:grid_example}%
		Two-dimensional ABC momentum grid. Red, solid circles represent
		spheres that are fully taken into account, while the blue, dashed
		one is excluded in our summations (see text for details).
	}%
\end{figure}

From a numerical point of view, it is beneficial to rearrange the three
summations in Eq.~\eqref{eq:integral_to_sum} such that they resemble
a spherical coordinate system \cite{Fischer:2002eq},
\begin{equation}
	\label{eq:spherical_summation}
	\frac{1}{L^{3}} \sum_{\vec{z} \in \bbZ^{3}} K(\vec{q}_{\vec{z}})
	=
	\frac{1}{L^{3}} \+ \sum_{j,\+ m} \+ K(\vec{q}_{jm}) \, .
\end{equation}
The index $j$ labels the spheres with constant radius
$\abs{\vec{q}_{\vec{z}}}$, and $m = m(j)$ denotes the multiplicity of the
individual momentum vectors on a given sphere $j$. The corresponding
vectors are denoted by $\vec{q}_{jm}$; see Fig.~\ref{fig:grid_example} for a
two-dimensional sketch.

Within the spherical momentum summation, it is straightforward to
implement an $\OO(3)$ invariant cutoff $\Lambda$. To this end, we distinguish
between ``complete'' and ``incomplete'' spheres: while the number of points on
complete spheres remains the same when the number of grid points in each
Cartesian direction is increased, the incomplete ones obtain additional points.
Thus, on each (sufficiently large) grid we have a large number of complete
spheres in the innermost region up to some distance from the origin, whereas
quite a large number of the outermost spheres are necessarily incomplete. We
sum only over complete spheres (red, solid circles in
Fig.~\ref{fig:grid_example}) and discard the incomplete ones (blue, dashed
circle in Fig.~\ref{fig:grid_example}). This procedure helps to reduce
so-called cubic artifacts \cite{Fischer:2002eq,Fischer:2005nf}. However, as
we will discuss in the next section, additional efforts are necessary to get
rid of these completely.

\subsubsection{\label{framework:prop_fv:improved_torus}%
	Finite volume with UV improvement
}

The spherical finite-volume summation introduced above has a severe drawback in
terms of numerical cost. The larger the grid the more dense are the complete
outer spheres and the more points are on every one of these. On the other hand,
since the dressing functions run logarithmically as functions of large squared
momenta in the UV, they do not change much from outer sphere to outer sphere.
Therefore, a lot of numerical effort is spent to integrate an almost constant
function. This effort can be drastically reduced by the following procedure.
We consider discrete spheres up to some matching cutoff $\Lambda_{\upvol}$ and
replace the spheres with radii between $\Lambda_{\upvol}^2 < q^2 < \Lambda^2$
with a continuous momentum integral. As a consequence, the original replacement
Eq.~\eqref{eq:integral_to_sum} is modified such that an additional integral
over the continuous momenta is added to the sum over the spatial modes,
\begin{align}
	\label{eq:integral_to_sum_improved}
	\int \frac{\dd^{3} q}{(2 \pi)^{3}} \, K(\vec{q})
	&\to
	\frac{1}{L^{3}}
	\sum_{\vec{z} \in \bbZ^{3}}^{\abs{\vec{q}_{\vec{z}}} < \Lambda_{\upvol}}
	K(\vec{q}_{\vec{z}})
	\notag
	\\[0.25em]
	&\phantom{\to}
	+ \int_{\abs{\vec{q}} > \Lambda_{\upvol}}
	\frac{\dd^{3} q}{(2 \pi)^{3}} \, K(\vec{q})
	\notag
	\\[1em]
	&\equiv
	\sumint_{\!\!\vec{q}} \+ K(\vec{q}) \, .
\end{align}

This modification, called ``UV improvement'' in the following, allows for much
larger values for the ultraviolet cutoff
$\Lambda$, which should mitigate artifacts that are caused by a potentially too
small cutoff. In addition, this improvement also allows to renormalize the
system at a large subtraction point where medium and volume effects are
negligible. Thus, the renormalization procedure can be carried out identically
as in infinite-volume calculations. Recently, a similar treatment has been used
in Ref.~\cite{Xu:2020loz} within a simpler truncation of the corresponding DSEs.

In practice, we observe that our results do not depend on $\Lambda_{\upvol}$ as
long as  $\Lambda_{\upvol}$ is larger than any other typical scale of the
system like temperature, chemical potential, and quark masses. This gives us
the freedom to choose different matching points for different box sizes $L$ and
essentially work with a fixed number of grid points in our three-dimensional
grid for momenta with magnitudes below $\Lambda_{\upvol}$.

In Sec.~\ref{results} below, we will discuss results from the unimproved
treatment of the system in the box as well as the UV improved one and
systematically compare both approaches.

\subsubsection{\label{framework:prop_fv:zero_mode}%
	Inclusion of the zero mode
}

In order to properly include the PBC zero modes of the external momentum $\vec{p}$
and the loop momentum $\vec{q}$ into our numerical setup, we have to make some
minor adjustments.
\begin{itemize}
\item[(i)] Our kernels are not well-defined at $\vec{q} = \vec{0}$, but smooth
in the  limit $\abs{\vec{q}} \to 0$. Therefore, we set the magnitude of the zero
mode  to a small but finite value, $\abs{\vec{q}_{\+\upzero}} = \varepsilon$,
introducing an effective infrared cutoff. We have checked explicitly that
variations of $\varepsilon$ over several orders of magnitude from
$\varepsilon=\SI{1}{\MeV}$ down to $\varepsilon = 10^{-7} \, \si{\MeV}$ lead to
no noticeable differences in our results.
\item[(ii)] The DSE kernels depend not only on the internal loop momentum
$\vec{q}$ but also on the external momentum $\vec{p}$ and their dot product,
which contains information on directions. The evaluation of this expression has
to be modified if either $\vec{p}$ or $\vec{q}$ corresponds to a zero mode. In
case of an internal zero mode, we set $\vec{p} \cdot \vec{q} = 0$. On the other
hand, in case of an external zero mode, the angular information contained in
$\vec{p} \cdot \vec{q}$ is important for the spherical sum over the loop
momentum $\vec{q}$ with multiplicity index $m$ in
Eq.~\eqref{eq:spherical_summation}. Therefore, in this case, we use
$\abs{\vec{p}_{\+\upzero}} = \varepsilon$ and employ the same directions/angles
for the zero-mode momenta as for the first finite momentum shell.
\end{itemize}

As already mentioned above, we refer to our setup with periodic boundary
conditions including the zero mode as PBC and compare later with results
with discarded zero mode denoted by $\PBCs$. Of course, the numerical effort
for PBC is increased compared to $\PBCs$. Thus, while we will show results
for $\PBCs$ in both setups---with and without UV improvement---we will provide
results for PBC only in the higher-quality UV-improved setup.

\subsection{\label{framework:dse}%
	Dyson--Schwinger equations
}

In order to obtain the dressed quark and gluon propagators,
Eqs.~\eqref{eq:quark_propagator} and \eqref{eq:gluon_propagator}, at nonzero
temperature and quark chemical potential, we solve a coupled set of truncated
DSEs in Landau gauge. Our setup takes the back reaction of the quarks onto the
Yang--Mills sector explicitly into account; for recent works (in infinite
volume) see, e.g., Refs.~\cite{Isserstedt:2019pgx,Gunkel:2020wcl,
Isserstedt:2020qll} as well as the comprehensive review \cite{Fischer:2018sdj}
and references therein.

Considering the introduction of finite volume as described in the previous
section, the dressed quark and gluon propagators are solutions of the DSEs
\begin{align}
	\label{eq:quark_DSE}
	S_{f}^{-1}(p)
	&=
	S_{0,\+f}^{-1}(p)
	+
	\Sigma_{f}(p) \, ,
	\\[0.5em]
	\label{eq:gluon_DSE}
	D_{\nu\sigma}^{-1}(p)
	&=
	\bigl[ D_{\nu\sigma}^{\textup{YM}}(p) \bigr]^{-1}
	+
	\Pi_{\nu\sigma}(p) \, ,
\end{align}
where $p = (\omega_n, \vec{p}_{il})$; $i$ and $l$ denote the sphere and
multiplicity, respectively, of the external momentum, which is discrete, too.
Here, $D_{\nu\sigma}^{\textup{YM}}$ denotes the sum of the inverse bare gluon
propagator and all diagrams with no explicit quark content while
\begin{equation}
	\label{eq:quark_propagator_bare}
	S_{0,\+f}^{-1}(p)
	=
	Z_{2}^{f} \+ \bigl( \ii \+ \omega_{n} \gamma_{4}
	+
	\ii \+ \vec{\gamma} \cdot \vec{p}_{il}
	+
	Z_{m_f} m_{f} \bigr)
\end{equation}
is the bare quark propagator with wave function renormalization constant
$Z_{2}^{f}$, mass renormalization constant $Z_{m_{f}}$, and renormalized
bare quark mass $m_{f}$. The quark self-energy $\Sigma_{f}$ and the quark-loop
contribution $\Pi_{\nu\sigma}$ to the gluon self-energy read
\begin{equation}
	\label{eq:quark_self_energy}
	\Sigma_{f}(p)
	=
	g^{2} \+
	\frac{4}{3}
	\frac{Z_{2}^{f}}{\tilde{Z}_{3}} \+
	T \+
	\sum_{\omega_{k}}
	\sumint_{\!\!\vec{q}} \+
	D_{\nu\sigma}(q - p) \+
	\gamma_{\nu} \+
	S_{f}(q) \+
	\Gamma_{\sigma}^{f}(q, p)
\end{equation}
and
\begin{equation}
\begin{aligned}
	\label{eq:gluon_quark_loop}
	\Pi_{\nu\sigma}(p)
	&=
	-\frac{g^{2}}{2} \+
	\sum_f \+
	\frac{Z_{2}^{f}}{\tilde{Z}_{3}} \+
	T \+
	\sum_{\omega_{k}}
	\sumint_{\!\!\vec{q}} \+
	\Tr \bigl[ \gamma_{\nu} \+ S_{f}(q)
	\\[0.25em]
	&\phantom{=\;}
	\times
	\Gamma_{\sigma}^{f}(q, q - p) \+
	S_f(q - p) \+ \bigr] \+ ,
\end{aligned}
\end{equation}
where $q = (\omega_k, \vec{q}_{jm})$ denotes the loop momentum. Furthermore,
$g$ is the strong coupling constant, $\tilde{Z}_{3}$ the ghost renormalization
constant, and $\Gamma_{\sigma}^{f}$ the dressed quark-gluon vertex. The
prefactors $4 \+ / \+ 3$ in Eq.~\eqref{eq:quark_self_energy} and $1 / \+ 2$
in Eq.~\eqref{eq:gluon_quark_loop} result from the color traces.

To get a closed system of equations, we proceed as follows. First, we replace
$D_{\nu\sigma}^{\textup{YM}}$ by quenched temperature-dependent lattice data
\cite{Fischer:2010fx,Maas:2011ez}. The explicit form of these fits can be
found in the Appendix of Ref.~\cite{Eichmann:2015kfa}. Second, we use an ansatz
for the dressed quark-gluon vertex: the leading Dirac tensor structure
$\gamma_{\sigma}$ of the Ball--Chiu vertex \cite{Ball:1980ay} is supplemented by
a phenomenological dressing function that accounts for non-Abelian effects and
guarantees the correct perturbative running of the propagators in the UV. The
explicit form of the vertex ansatz is detailed, e.g., in
Ref.~\cite{Fischer:2014ata}. Since this truncation is identical to the one used
in previous works, we refer the reader to Refs.~\cite{Fischer:2014ata,
Eichmann:2015kfa,Fischer:2018sdj,Isserstedt:2019pgx} and do not show more
detailed expressions in order to keep this section concise.

The set of truncated DSEs obtained in this way is solved self-consistently.
This gives us access to the nonperturbative quark and gluon propagators at
arbitrary temperature and chemical potential, i.e., across the phase diagram of
QCD. Finally, by varying $L$, we can investigate how the structure of the
latter changes in a finite volume. We use $2 + 1$ quark flavors that are
nontrivially coupled through the quark loop, which is evaluated explicitly. As
a consequence, the gluon becomes sensitive to the chiral dynamics of the quarks.

The phase structure is obtained by monitoring the behavior of the quark
condensate. It is the order parameter for chiral symmetry breaking and obtained
from the quark propagator according to
\begin{equation}
	\label{eq:condensate}
	\expval{\bar{\psi} \psi}_{f}
	=
	-3 \+ Z_{2}^{f} Z_{m_{f}} \+ T \+
	\sum_{\omega_{k}}
	\sumint_{\!\!\vec{q}} \+
	\Tr\bigl[ S_f(q) \bigr] \+ ,
\end{equation}
where the factor three stems from the color trace, and the remaining trace is
evaluated in Dirac space. The quark condensate is divergent for all flavors
with a nonzero bare mass because it contains a term proportional to
$m_{f} \+ \Lambda^{2}$. In our $(2 + 1)$-flavor setup, a regularized expression
can be obtained by considering the difference
\begin{equation}
	\label{eq:condensate_subtracted}
	\Delta_{\upu\ups}
	=
	\expval{\bar{\psi} \psi}_{\upu}
	-
	\frac{Z_{m_{\upu}}}{Z_{m_{\ups}}}
	\frac{m_{\upu}}{m_{\ups}} \+
	\expval{\bar{\psi} \psi}_{\ups}
	\, ,
\end{equation}
which defines the subtracted quark condensate. The divergent part of the
strange quark condensate cancels the corresponding one of the up quark
condensate when the former is multiplied with the up-to-strange mass ratio.
We use the inflection point of $\Delta_{\upu\ups}$ with temperature to define
the pseudocritical chiral transition temperature,
\begin{equation}
	\label{eq:Tc}
	\Tc
	=
	\argmax_{T}
	\left\rvert
	\frac{\partial \Delta_{\upu\ups}}{\partial \+ T}
	\right\rvert .
\end{equation}

We work in the isospin-symmetric limit of two degenerate light quarks
($m_{\upu} = m_{\upd}$, $\mu_{\upu} = \mu_{\upd}$) and choose $\mu_{\ups} = 0$.
The baryon chemical potential is then given by $\mu_{\upB} = 3 \+ \mu_{\upu}$.
We use an up/down quark mass of $m_{\upu} = m_{\upd} = \SI{0.8}{\MeV}$ at a
renormalization point of $\SI{80}{\GeV}$ (see Ref.~\cite{Isserstedt:2019pgx}
for details), and the mass ratio $m_{\ups} \+ / \+ m_{\upu} = 25.7$ is obtained
by applying the Bethe--Salpeter framework of Ref.~\cite{Heupel:2014ina} to our
truncation scheme and demanding physical pion and kaon masses in vacuum.

\section{\label{results}%
	Results and discussion
}

\begin{figure*}[t]
	\centering%
	\includegraphics[scale=1.0]{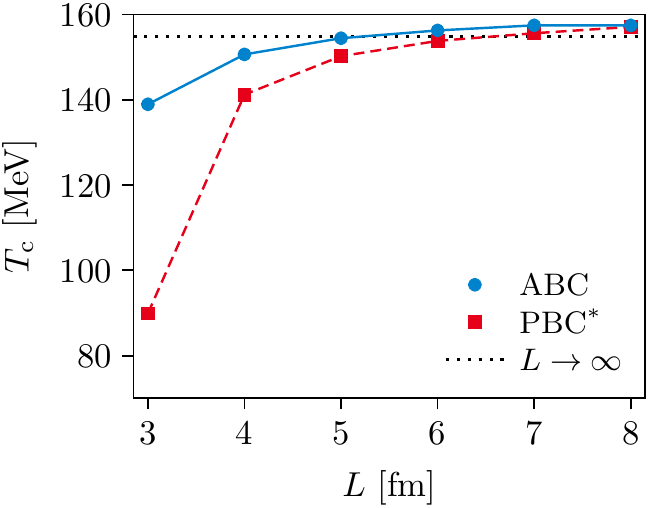}
	\hspace*{2em}
	\includegraphics[scale=1.0]{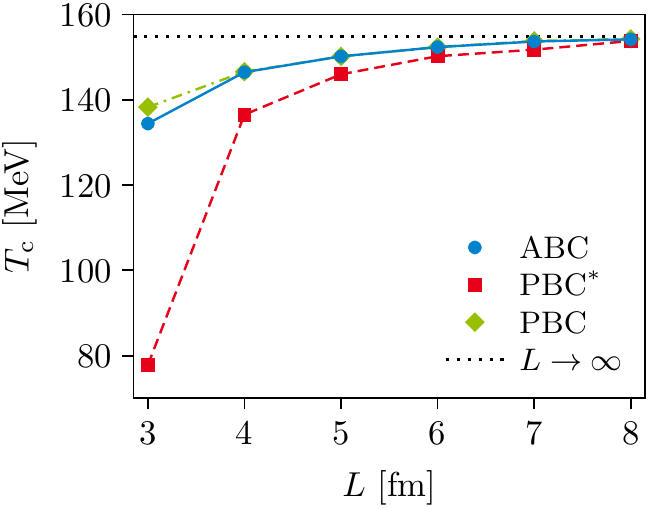}%
	\vspace*{-1mm}%
	\caption{\label{fig:Tc}%
		Finite-volume effects without (left) and with (right) UV improvement.
		We display the dependence of the pseudocritical chiral transition
		temperature on the box size for ABC (blue circles), $\PBCs$ (red
		squares), and PBC (green diamonds). The black, dotted line is the
		infinite-volume result. Data points are connected by lines to guide
		the eye.
	}%
\end{figure*}

Before we start with a discussion of our findings, let us remark how we
determine the pseudocritical chiral transition temperature. To damp numerical
instabilities in the condensates at finite volumes in temperature direction,
we employ a hyperbolic tangent fit that represents the condensate very well
up to chemical potentials around the CEP, i.e., in the crossover region. The
inflection point of the fit function determines $\Tc$. For the sake of
consistency, we also apply this fit procedure to the infinite-volume analysis.
This, in turn, leads to (very) small changes in the transition temperatures as
compared to previous works. For example, within the same truncation scheme we
find  $\Tc(L \to \infty) = \SI{155(1)}{\MeV}$ at zero chemical potential in this
work compared to $\Tc(L \to \infty) = \SI{156(1)}{\MeV}$ in the infinite-volume
calculation of Ref.~\cite{Isserstedt:2019pgx}. We emphasize that the difference
is purely technical and very small.

In the following, we discuss and compare our findings for finite-volume effects
without UV improvement with the ones including the UV improvement as discussed
in Sec.~\ref{framework:prop_fv:improved_torus}. We studied systems in boxes
with edge lengths of $L = 3$, $4$, $5$, $6$ and $\SI{8}{\fm}$ and compare to the
infinite-volume limit that is explicitly calculated using the framework of
Ref.~\cite{Isserstedt:2019pgx}.

\subsection{\label{results:Tc}%
	Pseudocritical chiral transition temperature at vanishing chemical potential
}

\begin{figure*}[t]
	\centering%
	\includegraphics[scale=1.0]{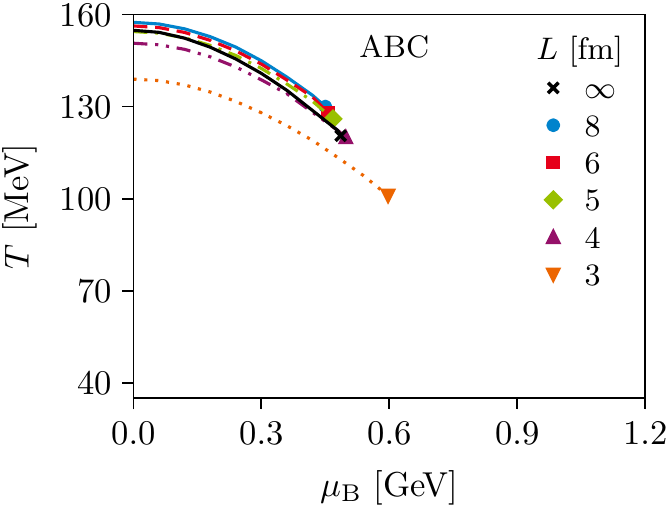}%
	\hspace*{2em}
	\includegraphics[scale=1.0]{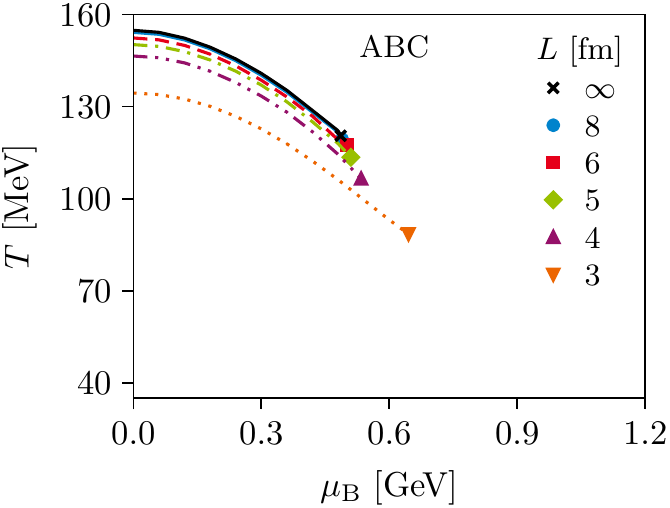}%
	\\[0.75em]%
	\includegraphics[scale=1.0]{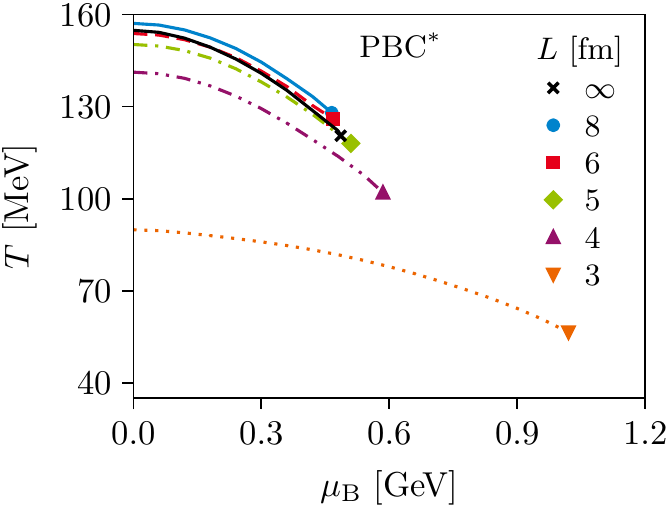}%
	\hspace*{2em}
	\includegraphics[scale=1.0]{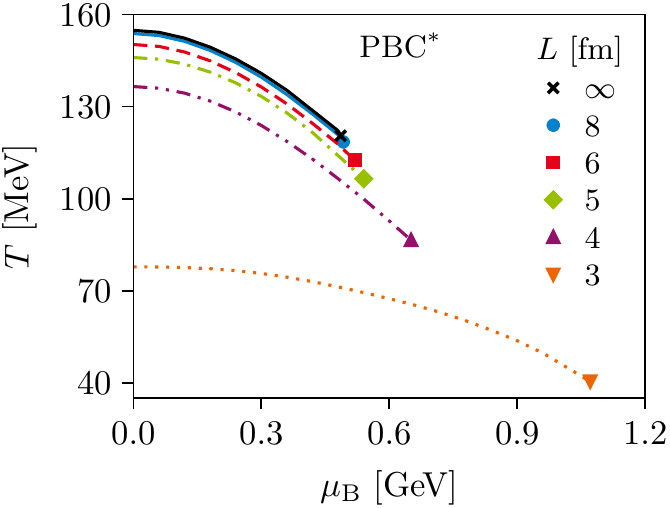}%
	\\[0.75em]%
	\hphantom{\includegraphics[scale=1.0]{CEP_PBC_pure.pdf}}%
	\hspace*{2em}
	\includegraphics[scale=1.0]{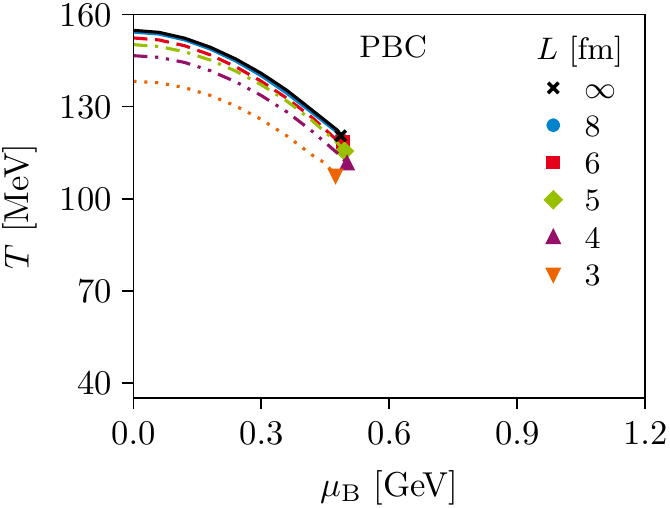}%
	\vspace*{-1mm}%
	\caption{\label{fig:CEP}%
		Finite-volume effects without (left) and with (right) UV improvement.
		Here, we show the crossover lines and locations of the CEPs (symbols)
		in the QCD phase diagram for different box sizes. The phase diagrams
		are obtained with ABC (upper row), $\PBCs$ (center row), and PBC (lower
		row).
	}%
\end{figure*}

In Fig.~\ref{fig:Tc}, we display the pseudocritical chiral transition
temperature $\Tc$ at vanishing chemical potential for antiperiodic and periodic
spatial boundary conditions with and without zero mode for the quarks.
For comparison, the infinite-volume result is indicated by a black,
dotted line. Comparing both figures, we clearly see the effect of the UV
improvement at large volumes. In the unimproved case, our results for both ABC
and $\PBCs$ suffer from cubic artifacts and overshoot the infinite-volume line. In
contrast, the improved results approach the infinite-volume results smoothly.\+%
\footnote{Technically, the lack of a consistent infinite-volume limit in the
unimproved calculations can be attributed to two reasons that are hard to
disentangle. First, the necessarily rather small UV cutoff, typically around
$\Lambda = \SI{10}{\GeV}$, in the case of the unimproved framework leads to
non-negligible cubic artifacts as already discussed above. In the improved
framework, the cutoff can be arbitrarily large and we work with $\Lambda =
\mathcal{O}(\SI{100}{\GeV})$. Second, the renormalization point in the
unimproved framework is necessarily located at even smaller momenta
($\SI{8}{\GeV}$ in our calculations) than the already small cutoff  and thus too
close to the infrared-momentum region, where medium and finite-volume effects
become important. As a consequence, the renormalization constants of the
unimproved calculation are contaminated by medium and finite-volume artifacts.
For the improved torus, this problem disappears because we are able to
renormalize in exactly the same fashion as in infinite-volume calculations,
i.e., with a renormalization point located deep inside the perturbative region.}
In this case, we also performed calculations at even larger box sizes, but since
the results are similar to the infinite-volume ones on the per mill level, we
did not include them in the plot.

At smaller box sizes, $\Tc$ decreases monotonously. While the decrease is
rather moderate down to $L \approx \SI{5}{\fm}$, volume effects become much
more pronounced for even smaller volumes. For example, at $L = \SI{3}{\fm}$
and $\PBCs$ we find that $\Tc$ is almost halved as compared to infinite volume.

In general, one can observe that quarks with $\PBCs$ are much more sensitive to
finite-volume effects across all investigated box sizes. This is caused by the
missing zero mode and the associated larger infrared cutoff introduced by the
discrete momentum grid.
From Eq.~\eqref{eq:matsubara_spatial}, the ratio of the smallest possible
momentum magnitude $p_{\upmin}$ between antiperiodic and periodic boundary
conditions is given by $p_{\upmin}^{\+\textup{ABC}} / \, p_{\upmin}^{\+\PBCs}
= \sqrt{3 \+ / \+ 4} \approx 0.866$.

From the right diagram of Fig.~\ref{fig:Tc}, it becomes apparent that the full
PBC results, i.e., with zero mode, resemble closely the ones for ABC. Down to
$L = \SI{4}{\fm}$, the ABC results lie on top of the PBC results with zero mode,
and they differ only by around three percent at our smallest investigated box
size of $L = \SI{3}{\fm}$.

For volumes as small as a box size of $L = \SI{3}{\fm}$, the system begins to
enter what is called the epsilon regime in chiral perturbation theory
\cite{Leutwyler:1992yt}. In this region, the product
$\alpha = m_f \+ V \+ \expval{\bar{\psi} \psi}_f^{L=\infty}$ of quark mass,
four-volume $V$, and infinite-volume quark condensate becomes of order one
and smaller and chiral symmetry starts to get restored already in the vacuum
theory. While the full effect (including critical scaling with $\alpha$) sets
in only at much smaller volumes ($L \lesssim \SI{2}{\fm}$), first effects are
already seen at our smallest box size of $L = \SI{3}{\fm}$.

\subsection{\label{results:phase_diagram}%
	Crossover line and critical endpoint
}

The finite-volume modifications on the phase structure are summarized in
Fig.~\ref{fig:CEP}. We show the phase diagrams for ABC (upper row), $\PBCs$
(center row), and PBC (lower row) of the quarks. For comparison, the
infinite-volume crossover line including the CEP is added, too. The left
diagrams correspond to results without UV improvement while the diagrams on the
right are obtained with UV improvement.

Similar to the results at zero chemical potential, we note the drastic effects
of the UV improvement. Whereas the CEPs of the series of larger and larger box
sizes do not approach the infinite-volume CEP without improvement (diagrams on
the left), they do so after the improvement has been implemented (diagrams on
the right). Both the crossover line and the CEP at $L = \SI{8}{\fm}$ are very
close to the infinite-volume limit. The remaining discrepancy is within the
numerical error of the infinite-volume calculation.\+%
\footnote{Note that the numerical error of the finite-volume calculations
is of the order of machine precision because no integrations are involved,
only sums.}
Furthermore, the volume-dependent shift of the crossover line and the CEP for
increasing box size approach the infinite-volume result uniformly. Thus, while
the overall qualitative behavior with and without improvement is the same,
quantitative aspects can only be discussed in the improved framework.

For ABC and $\PBCs$, both phase diagrams show a similar trend when the box size
is decreased: the CEP moves toward smaller temperatures and larger chemical
potentials. In more detail, we find that the increase of its location in
$\mu_{\upB}$ direction is larger than the decrease in $T$ direction leading to
a flattening of the chiral crossover line; see Sec.~\ref{results:curvature}
below where we discuss the volume dependence of the curvature of the crossover
line for different boundary conditions.

Down to $L = \SI{4}{\fm}$, the PBC results are generally very similar to the
corresponding ABC results. However, the volume-dependent movement of the CEP
is slightly slower and the endpoint values are closer to infinite volume. The
differences to ABC are around/below the ten-percent level, though. On the
other hand, at small volumes we observe a qualitative difference for full
PBC including the zero mode: the shift of the CEP in $\mu_{\upB}$ direction
inverts. In particular, the CEP in our smallest volume of
$L^3 = (\SI{3}{\fm})^3$ is located at smaller chemical potential than the
infinite-volume result. Nonetheless, the crossover lines again become flatter
for decreasing system sizes, which is in agreement with ABC and $\PBCs$.

Overall, the temperature dependence of the crossover line and CEP is analogous
to that of the pseudocritical chiral transition temperature at vanishing
chemical potential discussed above in Sec.~\ref{results:Tc}. Results for box sizes
$L \gtrsim \SI{5}{\fm}$ are rather similar to one another for the different
boundary conditions, while the phase diagrams for ABC, $\PBCs$, and PBC display
markedly different structures for small system sizes $L \lesssim \SI{4}{\fm}$.
Again, finite-volume effects are much more pronounced for $\PBCs$. Especially,
the result for $L = \SI{3}{\fm}$ stands out with a very flat crossover line and
a CEP at $(\mu_{\upB}, T) \approx (1070, 40) \, \si{\MeV}$ with UV improvement.

The volume dependence of the location of the CEP has been investigated also
in an FRG treatment within a two-flavor quark-meson model
truncation~\cite{Tripolt:2013zfa}. In this work, the position of the CEP has
been extracted from the maximum of the scalar susceptibility and its shift
with $L$ has been calculated for PBC between $L = \SI{4}{\fm}$ and
$L = \SI{10}{\fm}$. Compared to the DSE calculation presented here, the
infinite-volume CEP of the FRG analysis is generally located at (much) higher
chemical potential and lower temperature since its precise location depends on
the chosen infrared input parameters, in particular on the value of the
sigma-meson mass. However, a qualitative comparison of the results of
Ref.~\cite{Tripolt:2013zfa} with the ones presented here yields a satisfying
agreement between the present DSE and the FRG findings. Specifically, both the
$L$-dependent relative shift of the CEP as well as the onset of finite-volume
effects below $L = \SI{8}{\fm}$ coincide well. Below $L = \SI{4}{\fm}$, the CEP
disappeared completely in the FRG framework.

\subsection{\label{results:curvature}%
	Curvature of the chiral crossover line
}

Finally, we discuss the volume dependence of the curvature of the crossover
line. At small baryon chemical potential $\mu_{\upB}$, the crossover line can
be parameterized as
\begin{equation}
	\label{eq:curvature}
	\frac{\Tc(\mu_{\upB})}{\Tc}
	=
	1
	-
	\kappa_{2} \left( \frac{\mu_{\upB}}{\Tc} \right)^{2}
	-
	\kappa_{4} \left( \frac{\mu_{\upB}}{\Tc} \right)^{4}
	+
	\dotsm
	\, .
\end{equation}
Here, $\Tc(\mu_{\upB})$ and $\Tc = \Tc(0)$ are the pseudocritical chiral
transition temperatures at nonzero and vanishing chemical potential,
respectively, while the coefficient $\kappa_{2}$ is the curvature of the
transition line. We obtain $\kappa_{2}$ by fitting the values of the crossover
line at small baryon chemical potentials, $\mu_{\upB} \leq \SI{240}{\MeV}$, to
the parametrization given in Eq.~\eqref{eq:curvature}. For comparison, lattice
calculations yield a curvature in the range $0.0120 \leq \kappa_2 \leq 0.0153$
\cite{Bellwied:2015rza,HotQCD:2018pds,Bonati:2018nut,Borsanyi:2020fev}.

\begin{figure}[t]
	\centering%
	\includegraphics[scale=1.0]{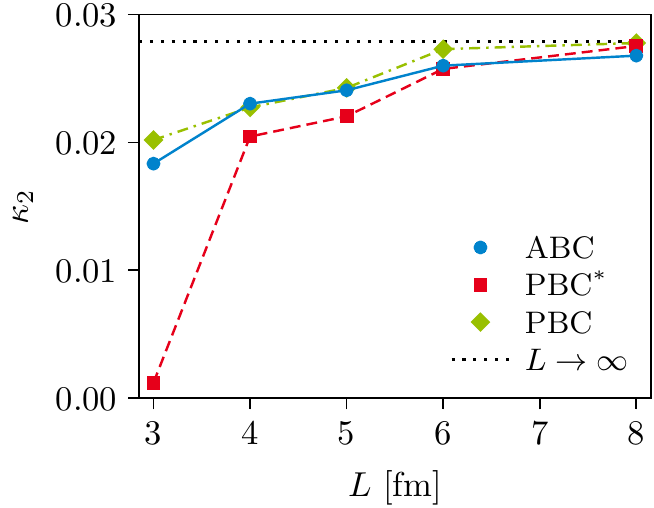}%
	\vspace*{-1mm}%
	\caption{\label{fig:kappa_improved}%
		UV-improved finite volume: $L$ dependence of the curvature of
		the crossover line for ABC (blue circles), $\PBCs$ (red squares), and
		PBC (green diamonds). Data points are connected by lines to guide the
		eye.
	}%
\end{figure}

Our results obtained with the UV improved framework are shown in
Fig.~\ref{fig:kappa_improved}. As already apparent from the phase diagrams
(see Fig.~\ref{fig:CEP}), the curvature is consistently smaller than the
infinite-volume result and decreases for smaller box sizes. Overall, this
flattening resembles the volume dependence of the pseudocritical chiral
transition temperature. That is, the results for $L = \SI{8}{\fm}$ are closest
to the infinite-volume value and drop monotonously with decreasing $L$ for all
boundary conditions.

Compared to the pseudocritical chiral transition temperatures shown in
Fig.~\ref{fig:Tc}, we find that the curvature displays a somewhat stronger
reaction to finite volume. Whereas the ABC temperatures are already close to
the infinite-volume result for $L = \SI{6}{\fm}$, the curvature parameter
$\kappa_2$ for both ABC and $\PBCs$ is still off by more than ten percent.
Only for very large box sizes of $L \gtrsim \SI{8}{\fm}$, we observe agreement
with the infinite-volume limit within errors. Here, it is important to note
that the fit is quite sensitive to details in the input data and choices of
fit intervals, such that $\kappa_2$ can only be extracted within a margin of
several percent.

In contrast to ABC and $\PBCs$, the PBC curvature is very close to the
infinite-volume result already above $L \gtrsim \SI{6}{\fm}$.
For $L = \SI{5}{\fm}$ and below, the PBC results are again very similar to the
ABC ones with slightly less pronounced finite-volume effects for
$L = \SI{3}{\fm}$.

The curvature for PBC between $L = \SI{2}{\fm}$ and $L = \SI{5}{\fm}$ was
studied with FRG techniques in Ref.~\cite{Braun:2011iz}. Above $L \approx
\SI{3}{\fm}$, there is qualitative agreement with our results: the curvature
increases with $L$. However, for smaller box sizes an interesting discrepancy
occurs. In our case, we find a monotonic decrease for smaller and smaller box
sizes, whereas the FRG results show an increase of the curvature when $L$ gets
smaller than $L \approx \SI{3.5}{\fm}$, resulting in an overall nonmonotonic
behavior. Even though this increase occurs for $L$ lower than we have
investigated here, the sharp drop of $\kappa_{2}$ for $L = \SI{3}{\fm}$ appears
to contradict such a scenario in our calculations. The reason for this
deviation is unknown. While the PBC zero mode is attributed to be the driving
force in the small-volume limit of Ref.~\cite{Braun:2011iz}, its inclusion does
not qualitatively change the behavior of the curvature at small $L$ in our case.
However, there are indications that these deviations might be rooted in
truncation/model artifacts, and we comment on this in the
\hyperref[appendix]{Appendix}.

\section{\label{summary}%
	Summary and conclusions
}

In this work, we studied the effects of a finite, uniform, three-dimensional
cubic volume with equal edge lengths $L$ and (anti)periodic boundary conditions
on the phase diagram of QCD. To this end, we employed a well-explored
truncation scheme for the coupled set of DSEs for the Landau gauge quark and
gluon propagators and extracted the volume dependence of the chiral order
parameter. In order to extract the volume behavior of the system, we found
two technical procedures to be mandatory: first, the removal of cubic artifacts
due to a UV improvement of the setup and, second, the explicit inclusion of
the zero mode for periodic boundary conditions.

For both types of boundary conditions, periodic and antiperiodic, we then find
similar and only moderate volume effects of the order of ten MeV and smaller
for box sizes $L \gtrsim \SI{5}{\fm}$. Only for very small volumes sizable
shifts of the CEP and the associated crossover line occur. These shifts are
almost monotonous: smaller volumes correspond to smaller transition
temperatures and the CEP shifts toward larger chemical potential. The only
deviation from this general behavior occurs for periodic boundary conditions
at very small box sizes. Our findings are consistent with corresponding
results from an FRG treatment of the quark-meson model.

In order to bridge the gap toward direct applications in the context of
heavy-ion collisions, (at least) two extensions of the current framework are
necessary. First, one needs to adapt the implementation of the finite-volume
boundary conditions to the actual physics case. There are no (anti)periodic
boundary conditions when two nuclei collide. The situation rather resembles
the one of a sphere (at large centrality) or an almond-shaped volume (at small
centrality) with a fuzzy boundary. This situation may be better represented in
a calculation in a sphere with MIT boundary conditions, as has been advocated
in Ref.~\cite{Xu:2020loz} within a DSE model framework. Second, one needs to
address the volume dependence of quantities such as fluctuations of conserved
charges \cite{Skokov:2012ds,Almasi:2016zqf}. Work in this direction within our
framework is in progress.


\begin{acknowledgments}
We thank Pascal J.~Gunkel and Richard Williams for discussions and Richard
Williams for comparisons at an early stage of this work. P.I.~is grateful for
the hospitality of the Nuclear Theory Group at the Lawrence Berkeley National
Laboratory where parts of this work were done. This work has been supported by
the Helmholtz Graduate School for Hadron and Ion Research for FAIR, the GSI
Helmholtzzentrum f\"{u}r Schwerionenforschung, and the BMBF under Contract
No.~05P18RGFCA.
\end{acknowledgments}

\appendix*
\section{\label{appendix}%
	Zero-mode contribution in different truncations
}

There have been discussions about the influence of the zero mode. In model
calculations, one can find that its inclusion leads to an increase of chiral
symmetry breaking for small box sizes $L \lesssim \SI{3}{\fm}$ compared to
large volumes, see, e.g., Refs.~\cite{Braun:2011iz,Almasi:2016zqf,
Juricic:2016tpt,Xu:2019gia}. In particular, the amount of chiral symmetry
breaking for small box sizes is found to be significantly larger than in
infinite volume.

In this work, however, we find that PBC with zero mode behaves qualitatively
and quantitatively almost identically to ABC. In order to investigate this
discrepancy, we also performed vacuum calculations for PBC with zero mode in
two simpler truncations: the quenched version of the truncation scheme
used in this work and one with a model gluon propagator. Both take no
backcoupling of quarks onto the gluon into account, i.e., the gluon can be seen
as a static input for the quark DSE, and unquenching effects are either absent
(quenched gluon) or modeled (model gluon). In more detail, the former
truncation is the one discussed in Sec.~\ref{framework:dse}, but the gluon in
the quark DSE is solely given by our fits $D_{\nu\sigma}^\textup{YM}$ to
quenched lattice data, while for the model gluon we resort to the
well-established Maris--Tandy model \cite{Maris:1999nt}. Furthermore, within
both truncations, we also varied the vertex ansatz by employing both the
Ball--Chiu-inspired vertex used in this work and the bare vertex
$\Gamma^{f (\textup{bare})}_{\sigma} = Z^{f}_{2} \+ \gamma_{\sigma}$.

\begin{figure}[t]
	\centering%
	\includegraphics[scale=1.0]{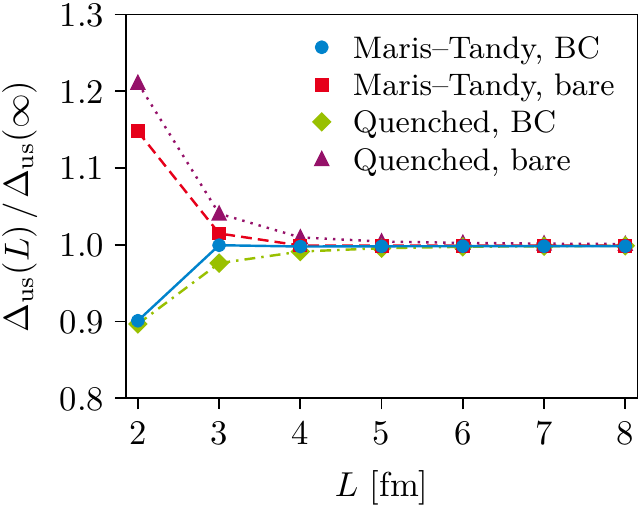}%
	\vspace*{-1mm}%
	\caption{\label{fig:appendix}%
		Truncations without quark backcoupling onto the gluon in vacuum:
		$L$ dependence of the subtracted quark condensate normalized to its
		infinite-volume value with a Ball--Chiu-inspired (BC) and a bare
		vertex. Data points are connected by lines to guide the eye.
	}%
\end{figure}

In Fig.~\ref{fig:appendix}, we show the $L$ dependence of the subtracted quark
condensate [Eq.~\eqref{eq:condensate_subtracted}] in vacuum normalized to its
infinite-volume value for both truncations with the two vertex ans\"{a}tze,
respectively. We investigated system sizes in the range of
$L = \numrange{2}{8} \, \si{\fm}$ since visible volume effects generally occur
at smaller $L$ for these truncations. In the large-$L$ limit, we notice that
all results tend to the infinite-volume value regardless of truncation and
vertex. For small $L$, however, especially below $L \lesssim \SI{3}{\fm}$, the
qualitative behavior depends crucially on the vertex ansatz. In case of the
bare vertex, we are able to reproduce the effects seen in model calculations:
an increasing condensate with decreasing system size in both truncations. In
case of the Ball--Chiu-inspired vertex, though, the $L$ dependence of the
condensate is in line with our findings in this work.

Therefore, we suppose that the zero-mode effects in the small-$L$ limit seen
in Refs.~\cite{Braun:2011iz,Almasi:2016zqf,Juricic:2016tpt,Xu:2019gia} are
either model artifacts or artifacts induced by approximations (e.g., mean field)
within the models.

\newpage

\urlstyle{same}
\bibliography{FiniteVolumeCEPBibliography}

\end{document}